%% file: main.tex
\documentclass[camera,letterpaper,nomarginnotes,nonarrowgutter]{jpaper}
\usepackage{mathptmx} 

\usepackage{fancyhdr}
\usepackage[normalem]{ulem}
\usepackage{microtype}
\usepackage{graphicx}
\usepackage{siunitx}
\usepackage{setspace}
\usepackage{flushend}
\usepackage{balance}
\usepackage{adjustbox}
\usepackage{ifthen}
\usepackage{algorithm2e}


\input{macros}
\input{glossary}

\usepackage{enumitem}
\usepackage{clipboard}

\usepackage[dvipsnames]{xcolor}
\ifcameraready
\else
\usepackage[colorinlistoftodos,prependcaption,textsize=small]{todonotes}
\fi
\usepackage{xargs} 

\usepackage{tikz}
\usepackage[most]{tcolorbox}
\tcbuselibrary{breakable}
\tcbuselibrary{hooks}

\usepackage{multirow}
\usepackage{datetime}
\usepackage{booktabs}
\usepackage{pifont}
\usepackage{hhline}

\ifcameraready
\else
\usepackage[colorinlistoftodos,prependcaption,textsize=small]{todonotes}
\fi
\usepackage{cite}
\usepackage[bookmarks=true,breaklinks=true,hidelinks]{hyperref}

\def\BibTeX{{\rm B\kern-.05em{\sc i\kern-.025em b}\kern-.08em
    T\kern-.1667em\lower.7ex\hbox{E}\kern-.125emX}}

\title{{An Experimental Analysis of RowHammer in HBM2 {DRAM} Chips}}

\author{\vspace{-18pt}\\%
\fontsize{11}{12}\selectfont%
{Ataberk Olgun$^{1}$}\quad%
{Majd Osseiran$^{1,2}$}\quad%
{A. Giray Ya\u{g}l{\i}k\c{c}{\i}$^{1}$}\quad%
{Yahya Can Tu\u{g}rul$^{1}$}\quad%
\\%
\fontsize{11}{12}\selectfont
{Haocong Luo$^{1}$}\quad
{Steve Rhyner$^{1}$}\quad
{Behzad Salami$^{1}$}\quad
{Juan Gomez Luna$^{1}$}\quad%
{Onur Mutlu$^{1}$}\quad
\vspace{0pt}\\%
{\fontsize{10}{11}\selectfont
$^1$\emph{SAFARI Research Group, ETH Z{\"u}rich}
\qquad
$^2$\emph{American University of Beirut}%
}
}

\begin{document}

\maketitle
\ifcameraready
    \fancypagestyle{firstpage}{
        \fancyhf{}
        \renewcommand{\headrulewidth}{0pt}
        \fancyhead[C]{\textcolor{blue}{DSN 2023 Disrupt -- Confidential Draft DO NOT SHARE! -- \today{}~\currenttime{}}} 
        \fancyfoot[C]{\thepage}
    }
    \thispagestyle{plain}
    \pagestyle{plain}
    \pagenumbering{arabic}
\else
    \fancypagestyle{firstpage}{
        \fancyhf{}
        \renewcommand{\headrulewidth}{0pt}
        \fancyhead[C]{\textcolor{blue}{DSN 2023 Disrupt -- Camera Ready -- Version \arabic{crversion}.1~~[\today{}~\currenttime{}]}}
        \fancyfoot[C]{\thepage}
    }
    \thispagestyle{firstpage}
    \pagestyle{plain}
    \pagenumbering{arabic}
\fi
\input{sections/00_abstract.tex}

\setstretch{0.972}
\input{sections/01_introduction.tex}
\input{sections/02_background.tex}
\input{sections/03_infrastructure.tex}

\input{sections/04_results.tex}
\input{sections/06_future_work.tex}
\input{sections/07_related_work.tex}
\input{sections/08_conclusion.tex}

\section*{Acknowledgments} 
We thank the anonymous reviewers of DSN 2023 Disrupt for feedback. We thank {the} SAFARI Research Group members for {valuable} feedback and the stimulating \omcr{2}{scientific and} intellectual environment they provide. We acknowledge the generous gift funding provided by our industrial partners ({especially} Google, Huawei, Intel, Microsoft, VMware), which has been instrumental in enabling the decade-long research we have been conducting on read disturbance in DRAM. This work was in part supported by the {Google Security and Privacy Research Award and the Microsoft Swiss Joint Research Center}.
\balance
\bibliographystyle{unsrt}
\bibliography{ref}
\pagebreak

\end{document}

%% file: macros.tex
\newif\ifdraft
\draftfalse

\newif\ifcameraready
\camerareadytrue

\newcounter{crversion}
\ifcameraready
\else
\fi

\newcommand{\dingOne}{\ding{182}}
\newcommand{\dingTwo}{\ding{183}}
\newcommand{\dingThree}{\ding{184}}
\newcommand{\dingFour}{\ding{185}}
\newcommand{\dingFive}{\ding{186}}
\newcommand{\dingSix}{\ding{187}}
\newcommand{\ignore}[1]{}

\definecolor{gfored}{rgb}{0.580, 0.050, 0.211}
\definecolor{ao}{rgb}{0.007, 0.520, 0.867}
\definecolor{moegi}{rgb}{0.357, 0.537, 0.188}
\definecolor{jl}{rgb}{1.0, 0.2, 0.8}
\definecolor{brown(web)}{rgb}{0.65, 0.16, 0.16}
\definecolor{bisque}{rgb}{1.0, 0.89, 0.77}
\definecolor{nbs}{rgb}{0.88, 0.07, 0.37}
\definecolor{yt}{rgb}{0.58, 0.44, 0.86}
\definecolor{iy}{rgb}{0.0, 0.26, 0.15}

\newcommand{\agycr}[2]{\ifnum#1=\value{crversion}\textcolor{blue}{#2}\else{#2}\fi}
\newcommand{\stecr}[2]{\ifnum#1=\value{crversion}\textcolor{magenta}{#2}\else{#2}\fi}
\newcommand{\omcr}[2]{\ifnum#1=\value{crversion}\textcolor{gfored}{#2}\else#2\fi}
\newcommand{\agycrtodo}[2]{\ifnum#1=\value{crversion}\todo[size=\scriptsize, linecolor=orange, bordercolor=orange, backgroundcolor=white]{\textcolor{blue}{TODO:~#2}}\else{}\fi}
\newcommand{\agycrcomment}[2]{\ifnum#1=\value{crversion}\todo[size=\scriptsize, linecolor=orange, bordercolor=orange, backgroundcolor=white]{\textcolor{blue}{Giray:~#2}}\else{}\fi}
\newcommand{\omcrcomment}[2]{\ifnum#1=\value{crversion}\todo[size=\scriptsize, linecolor=orange, bordercolor=orange, backgroundcolor=white]{\textcolor{gfored}{Onur:~#2}}\else{}\fi}

\ifdraft    
    \newcommand{\param}[1]{\textcolor{red}{\textbf{#1}}}
    \newcommand{\nrh}[0]{$HC_{first}$}
    
    \newcommand{\atatodo}[1]{\textcolor{red}{\textbf{TODO:}#1}}
    \newcommand{\atbnote}[1]{\textcolor{blue}{\textbf{NOTE:}#1}}
    \newcommand{\agycomment}[1]{\textcolor{gfored}{\textbf{[@gy:} #1\textbf{]}}}
    \definecolor{ao}{rgb}{0.007, 0.520, 0.867}
    
    \newcommand{\stecomment}[1]{\textcolor{purple}{\textbf{[@ste:} #1\textbf{]}}}
    
    \newcommand{\atbcomment}[1]{\textcolor{ao}{\textbf{[@atb:} #1\textbf{]}}}
    \newcommand{\atbcommentside}[1]{\todo[size=\scriptsize, linecolor=orange, bordercolor=orange, backgroundcolor=white]{\textcolor{ao}{\textbf{@atb:} #1}}}
    
    \newcommand{\yctcomment}[1]{\textcolor{purple}{\textbf{[@yct:} #1\textbf{]}}}
    
    \newcommand{\majdcomment}[1]{\textcolor{mo}{\textbf{[@majd:} #1\textbf{]}}}
    
    \definecolor{mo}{rgb}{0, 0.5, 0}
    \newcommand{\majdtodo}[1]{\textcolor{red}{\textbf{TODO:}#1}}

    \newcommand{\om}[1]{\textcolor{blue}{#1}}    
    \newcommand{\omcomment}[1]{\todo[size=\scriptsize, linecolor=orange, bordercolor=orange, backgroundcolor=white]{\textcolor{blue}{\textbf{@onur:} #1}}}
    \newcommand{\omcommentinline}[1]{\textcolor{blue}{\textbf{[@Onur:} #1\textbf{]}}}

    \newcommand{\bscomment}[1]{\todo[size=\scriptsize, linecolor=brown, bordercolor=brown, backgroundcolor=white]{\textcolor{blue}{\textbf{@Behzad:} #1}}}
    \newcommand{\bscommentinline}[1]{\textcolor{brown}{\textbf{[@Behzad:} #1\textbf{]}}}
\else

    \newcommand{\om}[1]{{#1}}
    \newcommand{\omcomment}[1]{}
    \newcommand{\omcommentinline}[1]{}

    \newcommand{\nrh}[0]{$HC_{first}$}

    \newcommand{\bscomment}[1]{}
    \newcommand{\bscommentinline}[1]{}

    \newcommand{\agycomment}[1]{}

    \newcommand{\stecomment}[1]{}

    \newcommand{\atbcomment}[1]{}
    \newcommand{\atbcommentside}[1]{}
    \newcommand{\atatodo}[1]{}
    \newcommand{\atbnote}[1]{}

    \newcommand{\majdcomment}[1]{}
    \newcommand{\majdtodo}[1]{}

    \newcommand{\yctcomment}[1]{}

    \newcommand{\param}[1]{{#1}} 
\fi

\def\UrlBreaks{\do\/\do-\/\do.\/\do:}

\expandafter\def\expandafter\UrlBreaks\expandafter{\UrlBreaks
  \do\a\do\b\do\c\do\d\do\e\do\f\do\g\do\h\do\i\do\j
  \do\k\do\l\do\m\do\n\do\o\do\p\do\q\do\r\do\s\do\t
  \do\u\do\v\do\w\do\x\do\y\do\z\do\A\do\B\do\C\do\D
  \do\E\do\F\do\G\do\H\do\I\do\J\do\K\do\L\do\M\do\N
  \do\O\do\P\do\Q\do\R\do\S\do\T\do\U\do\V\do\W\do\X
  \do\Y\do\Z}

\newcommand{\exploitingRowHammerAllCitations}[0]{\cite{fournaris2017exploiting, poddebniak2018attacking, tatar2018throwhammer, carre2018openssl, barenghi2018software, zhang2018triggering, bhattacharya2018advanced, google-project-zero, kim2014flipping, rowhammergithub, seaborn2015exploiting, van2016drammer, gruss2016rowhammer, razavi2016flip, pessl2016drama, xiao2016one, bosman2016dedup, bhattacharya2016curious, burleson2016invited, qiao2016new, brasser2017can, jang2017sgx, aga2017good, mutlu2017rowhammer, tatar2018defeating, gruss2018another, lipp2018nethammer, van2018guardion, frigo2018grand, cojocar2019eccploit,  ji2019pinpoint, mutlu2019rowhammer, hong2019terminal, kwong2020rambleed, frigo2020trrespass, cojocar2020rowhammer, weissman2020jackhammer, zhang2020pthammer, yao2020deephammer, deridder2021smash, hassan2021utrr, jattke2022blacksmith, tol2022toward, kogler2022half, orosa2022spyhammer, zhang2022implicit, liu2022generating, cohen2022hammerscope, zheng2022trojvit, fahr2022frodo, tobah2022spechammer, rakin2022deepsteal, aydin2022cyber, mus2022jolt, wang2022research, lefforge2023reverse,fahr2022effects, kaur2022work, cai2022feasibility, li2022cyberradar, roohi2022efficient, staudigl2022neurohammer, yang2022socially, islam2022signature}}

\newcommand{\understandingRowHammerAllCitations}[0]{\cite{kim2014flipping, park2016statistical, park2016experiments,lim2017active, ryu2017overcoming, lim2018study, yun2018study, yang2019trap, walker2021ondramrowhammer, kim2020revisiting, orosa2021deeper, orosa2022spyhammer, cohen2022hammerscope, yaglikci2022understanding, khan2018analysis, agarwal2018rowhammer, li2014write, ni2018write, genssler2022reliability, mutlu2023fundamentally, he2023whistleblower, baeg2022estimation, frigo2020trrespass, mutlu2017rowhammer, mutlu2018rowhammer, mutlu2019rowhammer}}

\newcommand{\rowHammerGetsWorseCitations}[0]{\cite{kim2020revisiting, frigo2020trrespass, yaglikci2022understanding, orosa2021deeper, mutlu2017rowhammer, mutlu2018rowhammer, mutlu2019rowhammer, mutlu2023fundamentally}}

\newcounter{obs}
\newcounter{take}
\newcommand{\figref}[1]{Fig.~\ref{#1}}

\newcommand{\secref}[1]{§\ref{#1}}

\newcommand{\fnref}[1]{{\textsuperscript{\ref{#1}}}}

%% file: glossary.tex
\usepackage{glossaries}

\newacronym{vdd}{$V_{DD}$}{supply voltage}
\newacronym{vpp}{$V_{PP}$}{wordline voltage}
\newacronym{vppmin}{$V_{PPmin}$}{the lowest \gls{vpp} at which the DRAM module can successfully communicate with the FPGA}
\newacronym{vwl}{$V_{PP}$}{wordline voltage}
\newacronym{gnd}{$GND$}{ground}
\newacronym{hcfirst}{$HC_{first}$}{the minimum aggressor row activation count necessary to cause a RowHammer bitflip}
\newacronym{rblast}{$r_{Blast}$}{blast radius}
\newacronym{ber}{$BER$}{the fraction of DRAM cells that experience a bitflip in a DRAM row}
\newacronym{nhc}{$N_{HC}$}{hammer count}
\newacronym{hc}{$HC$}{hammer count}
\newacronym{trcd}{$t_{RCD}$}{row activation latency}
\newacronym{tcl}{$t_{CL}$}{column access latency}
\newacronym{tcwl}{$t_{CWL}$}{column write latency}
\newacronym{trp}{$t_{RP}$}{precharge latency}
\newacronym{trcdmin}{$t_{RCDmin}$}{{the minimum time delay required}}
\newacronym{tras}{$t_{RAS}$}{charge restoration latency}
\newacronym{trasmin}{$t_{RASmin}$}{the minimum latency required}
\newacronym{trefw}{$t_{REFW}$}{refresh window}
\newacronym{vgs}{$V_{GS}$}{gate-to-source voltage}
\newacronym{vthresh}{$V_{TH}$}{the voltage threshold that the bitline voltage should exceed for the activation to be reliably completed}
\newacronym{kde}{KDE}{kernel density estimate}
\newacronym{ref}{refresh}{$REF$}
\newacronym{tsv}{through-silicon-via}{TSV}
\newacronym{iqr}{IQR}{inter-quartile-range}
\newacronym{cov}{CV}{coefficient of variation}
\newacronym{wcdp}{\emph{WCDP}}{the worst-case data pattern}

%% file: sections/00_abstract.tex
\begin{abstract}

RowHammer (RH) is a significant and worsening security, \omcr{2}{safety,} and reliability issue of modern DRAM chips that can be exploited to break memory isolation. Therefore, it is important to understand real DRAM chips' {RH} characteristics. Unfortunately, no prior work extensively studies the {RH} vulnerability of modern 3D-stacked high-bandwidth memory {(HBM)} chips{, which are commonly used in modern GPUs}.

In this work, we experimentally characterize the {RH} vulnerability of a real HBM2 DRAM chip. We show that 1)~different 3D-stacked channels of HBM2 memory exhibit significantly different levels of {RH} vulnerability {(up to \SI{79}{\percent} difference in bit error rate)}, 2)~the DRAM rows 
at the end of a DRAM bank {(rows with the highest addresses)} exhibit significantly \agycr{1}{fewer RH bitflips} than \agycr{1}{other} rows,
and 3)~{a} modern HBM2 DRAM {chip} implement{s} {undisclosed} {RH} defenses that {are triggered by} periodic refresh operations.
{We describe the implications of our observations on future {RH} attacks and defenses \agycr{1}{and} discuss \agycr{1}{future work for} understanding {RH} in 3D-stacked memories.}

\end{abstract}

%% file: sections/01_introduction.tex
\glsunset{ber}
\glsunset{hcfirst}
\section{Introduction}
\label{sec:introduction}
{Modern DRAM chips} {{suffer from} the RowHammer (RH) phenomenon~\cite{kim2014flipping,mutlu2017rowhammer,mutlu2019retrospective,kim2020revisiting, mutlu2023fundamentally}} where repeatedly opening \omcr{2}{(i.e., activating)} and closing a DRAM row (i.e., aggressor row) induces bitflips in physically nearby rows (i.e., victim rows), breaking memory isolation~\agycr{2}{\understandingRowHammerAllCitations{}}.
{N}umerous studies {experimentally demonstrate that} a malicious attacker can {reliably} induce {{RH} bitflips} in a targeted manner to compromise system integrity, confidentiality, and availability~\exploitingRowHammerAllCitations{}. 
{\om{RH worsens} {in newer DRAM chips with {smaller technology nodes}, \om{where}} {RH} bitflips 1)~\omcr{2}{occur} with fewer row activations, e.g., more than $10\times$ reduction in less than a decade~\cite{kim2020revisiting}, and 2)~\om{manifest} {in} more DRAM cells, compared to older DRAM {chips}~\rowHammerGetsWorseCitations{}.}

{To meet the high bandwidth requirements of modern data-intensive applications \om{(e.g., GPU workloads~\cite{bakhoda2009analyzing,che2009rodinia, gomez2017chai, ghose2019demystifying})}, DRAM designers develop High Bandwidth Memory (HB{M})~\cite{jedec2015hbm} DRAM chips, which contain multiple layers of 3D-stacked DRAM dies, using cutting-edge technology nodes.}
{It is important to understand {RH} in HBM DRAM chips that have new architectural characteristics (e.g., multiple layers of DRAM dies and area- and energy-intensive through-silicon vias) which might \omcr{2}{potentially} affect the \agycr{2}{RH} vulnerability in various ways.}
{Such understanding can help identify potential {RH}-induced security\omcr{2}{, safety,} and reliability issues in HBM-based systems and allow system designers to develop effective and efficient defense mechanisms.}
{Unfortunately, \emph{no} prior work \omcr{2}{studies} the {RH} vulnerability of modern HBM DRAM chips.}

{\textbf{Our goal} in this work is to experimentally \om{analyze} how vulnerable HBM DRAM chips are to \stecr{2}{RH}.}
To this end, we provide {the first detailed} experimental characterization {of the {RH} vulnerability} in a {modern} HBM2 DRAM chip. We {provide} \param{two} main analyses in our study. 
{First, we analyze the spatial variation in \stecr{2}{RH} vulnerability based on the physical location of victim rows in terms of two metrics: the fraction of DRAM cells that experience a bitflip in a DRAM row (\omcr{2}{i.e.,} \omcr{3}{Bit Error Rate,} \gls{ber}) and the minimum aggressor row activation count necessary to cause a\agycr{2}{n} \stecr{2}{RH} bitflip (\omcr{2}{i.e.,} \gls{hcfirst}). {\secref{sec:spatial-variation-analysis} shows detailed results on how RH vulnerability \agycr{1}{varies} across HBM2 channels, pseudo channels, banks, and rows.}}
Second, we investigate undisclosed in-DRAM {RH} mitigations, triggered by periodic \agycr{2}{refreshes} (e.g., TRR~\cite{micron2016trr,frigo2020trrespass,hassan2021utrr}) in HBM2 {(\secref{sec:uncovering})}.\footnote{{The HBM2 standard~\cite{jedec2015hbm} specifies a \agycr{2}{mode called} Target Row Refresh (TRR). To enable TRR Mode, the memory controller issues a well-defined series of commands. \agycr{1}{Different} from \agycr{1}{this mode}, we investigate whether an \omcr{2}{\emph{undisclosed}} TRR mechanism \agycr{1}{is implemented} in \agycr{1}{the tested} chip.}}

{We summarize the \param{three} \omcr{2}{major} observations from our experimental analyses. First, different 3D-stacked channels of the HBM2 chip exhibit significantly different levels of RH vulnerability in \gls{ber} (up to \param{79\%}) and \gls{hcfirst} (up to \param{20\%}). Second, DRAM rows near the end of a DRAM bank (the last \param{832} \agycr{2}{out of 16K} rows) exhibit substantially smaller \gls{ber} than other DRAM rows. Third, the \agycr{1}{tested} HBM2 DRAM chip implements an \agycr{1}{undisclosed} in-DRAM \agycr{1}{RH} defense mechanism (\secref{sec:uncovering}). Our {experimental analyses} \omcr{2}{also} show that the RH vulnerability of a cell depends on i) the cell's physical location within a DRAM bank and ii) data stored in the neighboring cells, similar to prior works' findings in DDR3/4 DRAM chips~\cite{kim2014flipping,kim2020revisiting,orosa2021deeper}.}

{We make the following contributions:}
\begin{itemize}
    \item We present the first detailed experimental characterization of the \omcr{2}{RowHammer (RH)} vulnerability in a \agycr{1}{modern} HBM2 DRAM chip \agycr{1}{and} show \omcr{2}{that it is} susceptible to \agycr{1}{RH} bitflips.
    \item We show that the \agycr{1}{RH} vulnerability \agycr{1}{significantly} varies across \agycr{1}{HBM2 DRAM} channels and rows \agycr{1}{within} \om{each} channel. 
    \item We show that an HBM2 DRAM chip implements an undisclosed in-DRAM \agycr{1}{RH} mitigation \omcr{2}{mechanism, which} resembles the \agycr{2}{one} found in recent DDR4 DRAM chips manufactured by a major DRAM manufacturer~\cite{hassan2021utrr}.
\end{itemize}

%% file: sections/02_background.tex
\section{Background}

\subsection{HBM2 Organization \& Operation}

\begin{figure}[!h]
    \centering
    \includegraphics[width=\linewidth]{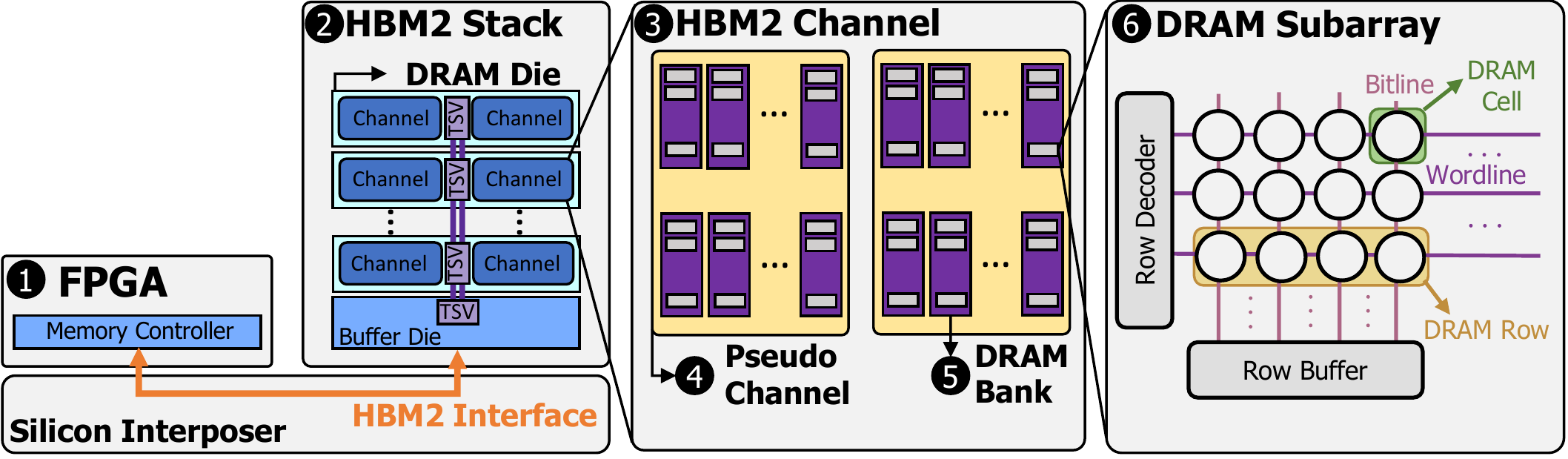}
    \caption{HBM2 \omcr{2}{DRAM System} Organization}
    \label{fig:hbm-organization}
\end{figure}

\figref{fig:hbm-organization} presents the organization of an HBM2 DRAM chip~\cite{jedec2015hbm, oconnor2021thesis} when used in an FPGA-based system. The \agycr{2}{FPGA (\dingOne{}) has a} memory controller \agycr{2}{communicating} with one or multiple stacks of HBM using the HBM2 interface \agycr{2}{via a silicon interposer}. An HBM2 stack \omcr{2}{(\dingTwo{})} contains multiple DRAM dies \agycr{1}{that are} stacked on top of the buffer die and connected using through-silicon vias (TSVs). Each HBM2 die \agycr{1}{has} one or \agycr{2}{more (shown two)} \agycr{1}{independent HBM2} channels. \omcr{2}{An HBM2 channel (\dingThree{}) consists of two pseudo channels (\dingFour{}), each of which with} multiple \agycr{2}{(e.g., 16)} DRAM banks.
A DRAM bank \omcr{2}{(\dingFive{})} \agycr{1}{consists of \omcr{3}{hundreds of} subarrays~\cite{salp,seshadri2013rowclone,chang2014improving}}. \agycr{2}{A DRAM subarray (\dingSix{})} has many DRAM cells that are laid in a two-dimensional array of rows and columns, \agycr{1}{and} 
a row buffer. \agycr{1}{DRAM cells are internally accessed \omcr{2}{at} DRAM row granularity by asserting a wire called wordline. Each DRAM cell in a row is connected to the row buffer \omcr{2}{(i.e., sense amplifiers) via} another wire called bitline.}

\noindent
\textbf{Operation.} 
{The memory controller issues an activate (ACT) command targeting a DRAM row to access a DRAM cell. The row decoder asserts the row's wordline, copying the data in the row to the row buffer.
To access a different cell in another DRAM row, the memory controller issues a precharge ($PRE$) command, which deasserts the wordline.}

\noindent
\textbf{Periodic refresh.} {A DRAM cell stores data as charge in its capacitor. \agycr{1}{Because} the capacitor loses charge over time, \agycr{1}{it} must be periodically refreshed to prevent data corruption. To refresh DRAM cells, the memory controller periodically issues refresh ($REF$) commands (e.g., every \SI{3.9}{\micro\second}) such that each cell is refreshed once at a fixed refresh \omcr{2}{period} (e.g., \SI{32}{\milli\second}).}

%% file: sections/03_infrastructure.tex
\section{Experimental Infrastructure}

We experimentally study an HBM2 chip using a modified version of the DRAM Bender testing infrastructure~\cite{olgun2022drambender,safari-drambender, hassan2017softmc, softmcgithub}. This infrastructure allows us to precisely control the HBM2 command timings at the granularity of \SI{1.66}{\nano\second} (i.e., the HBM2 interface clock speed is \param{600} MHz). 
{Our HBM2 chip has i) a stack density of 4 GiB, ii) 8 channels, iii) 2 pseudo channels, iv) 16 banks, v) 16384 rows, and vi) 32 columns.}

\begin{figure}[ht]
    \centering
    \includegraphics[width=0.8\linewidth]{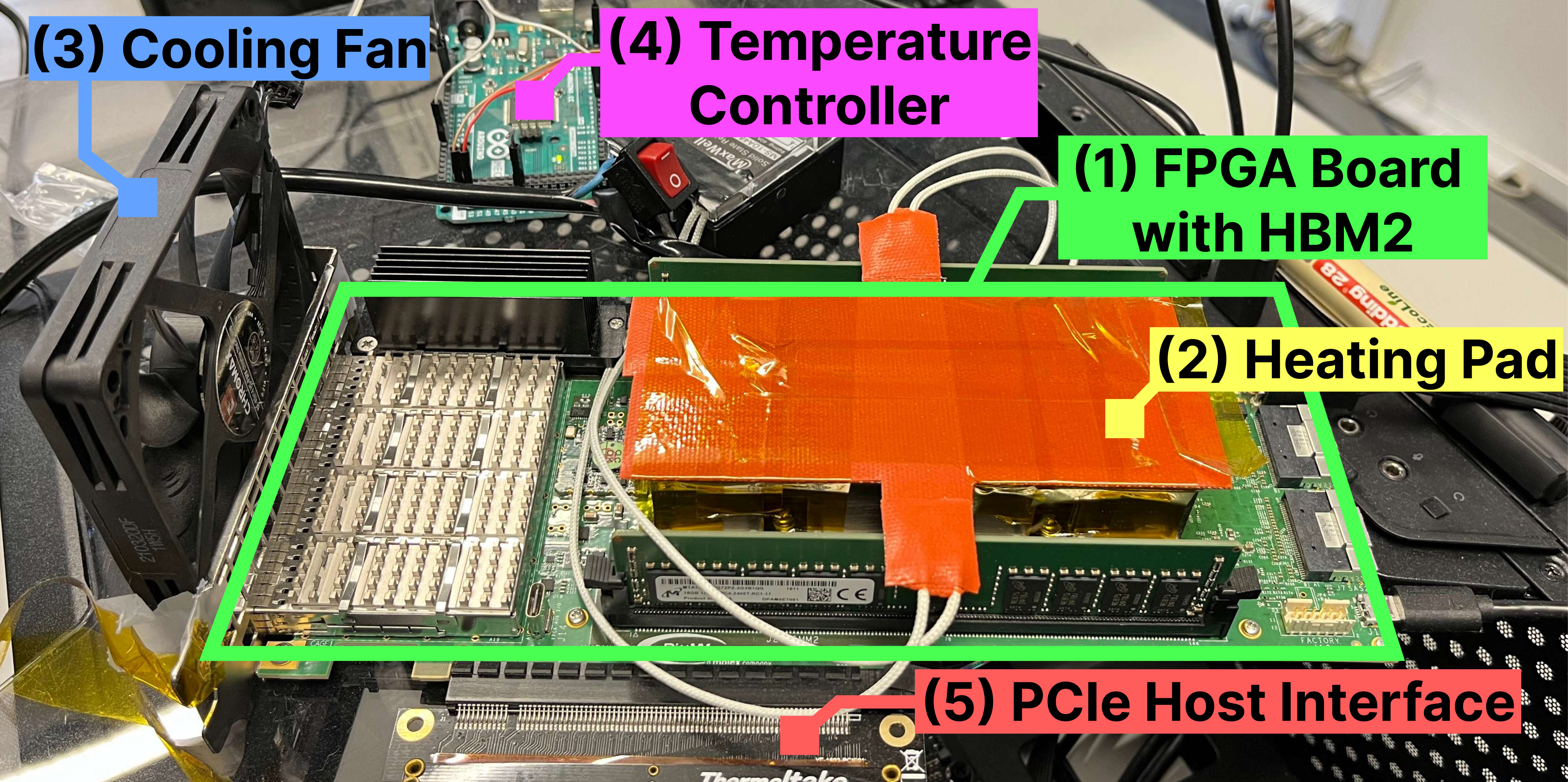}
    \caption{HBM2 DRAM testing infrastructure}
    \label{fig:dram-bender}
\end{figure}

\noindent
\textbf{Testing setup.}
Figure~\ref{fig:dram-bender} shows our testing setup. We conduct our experiments using a Bittware XUPVVH HBM2 FPGA board~\cite{xupvvh} (1). 
We use the heating pad (2) and the cooling fan (3) to \omcr{2}{change} the \omcr{2}{ambient}\agycrcomment{2}{I think it is ok to say ambient here because we heat up or cool down the heatsink} temperature of the HBM2 chip. The Arduino MEGA~\cite{arduinomega} temperature controller (4) communicates with i)~the host machine to retrieve a target temperature and ii)~the FPGA board to retrieve the HBM2 chip's temperature. The temperature controller controls the heating pad and the cooling fan using a closed-loop PID controller. A host machine executes the test programs described in \secref{sec:testing-methodology} on the FPGA board using the PCIe connection (5).

\subsection{Testing Methodology}
\label{sec:testing-methodology}

\noindent
\textbf{Disabling Sources of Interference}. 
{We identify \param{four} sources that can interfere with our characterization results: 1) periodic refresh~\cite{jedec2015hbm}, 2) on-die RH defense mechanisms (e.g., TRR~\cite{frigo2020trrespass,hassan2021utrr,micron2016trr}), 3) \agycr{2}{data} retention failures \cite{liu2013experimental, khan2014efficacy, khan2016parbor, qureshi2015avatar}, and 4) ECC \cite{patel2017reaper, patel2019understanding, patel2020beer, patel2021harp}. First, we do \emph{not} issue periodic refresh commands in our experiments. Second, disabling periodic refresh disables all known on-die RH defense mechanisms~\cite{orosa2021deeper,yaglikci2022understanding,kim2020revisiting,hassan2021utrr, luo2023rowpress}. Third, we ensure that our experiments finish within \agycr{2}{\SI{27}{\milli\second}, which is significantly smaller than the standard refresh period of \SI{32}{\milli\second}} where manufacturers guarantee no \agycr{2}{data} retention errors will occur~\cite{jedec2015hbm}. Fourth, we disable ECC by setting the corresponding HBM2 mode register bit to zero~\cite{jedec2015hbm}.}

\noindent
\textbf{RH Access Pattern}. 
We use double-sided RH~\cite{kim2014flipping,kim2020revisiting,orosa2021deeper,seaborn2015exploiting}, whereby we alternate \omcr{2}{activations to} each of the aggressor rows neighboring a victim row. 
{\agycr{1}{To find which rows are physically adjacent, we reverse-engineer the} logical (memory-controller-visible) to physical \omcr{2}{(in-DRAM)} row address \agycr{1}{mapping}~\cite{kim2014flipping, smith1981laser, horiguchi1997redundancy, keeth2001dram, itoh2013vlsi, liu2013experimental,seshadri2015gather, khan2016parbor, khan2017detecting, lee2017design, tatar2018defeating, barenghi2018software, cojocar2020rowhammer,  patel2020beer, yaglikci2021blockhammer, orosa2021deeper},}
{following the methodology described in prior work~\cite{orosa2021deeper}}.

\noindent
\textbf{RH Test Parameters}.  
We define one hammer as a pair of activations to the two aggressor rows. {{We measure two metrics in our tests\omcr{2}{:} $BER$ and $HC_{first}$, as defined in~\secref{sec:introduction}.}}
We use 256K hammers (i.e., 512K activations) in our $BER$ experiments and up to 256K hammers 
in our $HC_{first}$ experiments. {We repeat both experiments} for each of the four data patterns shown in Table~\ref{table_data_patterns}. We define \agycr{2}{\gls{wcdp}} as the data pattern that causes the smallest \gls{hcfirst} \omcr{2}{for a given row}. {When multiple} data patterns cause the \omcr{2}{smallest} \gls{hcfirst}, we select \gls{wcdp} as the data pattern that causes the largest \gls{ber} at a hammer count of 256K. To {maintain a reasonable experiment time}, we study the effects of RH on the first, middle, and last 3K rows {in} a bank {in every channel} and repeat all {experiments} five times. {The HBM2 chip's temperature is kept at} \agycr{2}{$85$°C, the maximum operating temperature at the nominal refresh rate,} {in all of our experiments}.\agycrcomment{3}{I removed $\pm1$ to avoid the concern of having data retention errors at 86C}

\begin{table}[htbp]
\caption{Data patterns used in our RH tests}
\vspace{-5mm}
\begin{center}
\begin{adjustbox}{max width=\linewidth}
\begin{tabular}{l||cccc}
\textbf{Row Addresses} & \textbf{\textit{Rowstripe0}}& \textbf{\textit{Rowstripe1}}& \textbf{\textit{Checkered0}} & \textbf{\textit{Checkered1}}\\
\hline
\hline
Victim (V) & 0x00 & 0xFF & 0x55 & 0xAA\\
Aggressors (V $\pm$ 1) & 0xFF & 0x00 & 0xAA & 0x55\\
V $\pm$ [2:8] & 0x00 & 0xFF & 0x55 & 0xAA\\
\end{tabular}
\end{adjustbox}
\label{table_data_patterns}
\vspace{-2em}
\end{center}
\end{table}

%% file: sections/04_results.tex
\section{Spatial Variation Analysis}
\label{sec:spatial-variation-analysis}

We provide the first spatial variation analysis of RH across HBM2 channels, pseudo channels, banks, and rows.

\begin{figure}[ht]
    \centering
    \includegraphics[width=\linewidth]{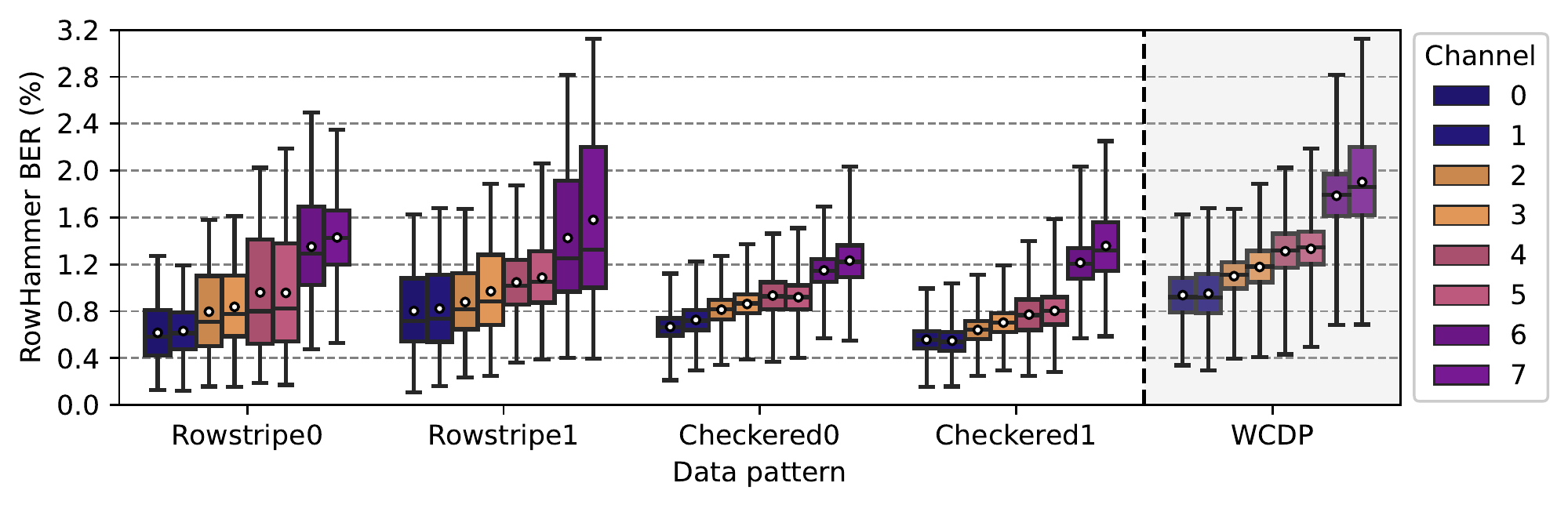}
    \vspace{-1em}
    \caption{\Glsfirst{ber} across different DRAM rows, channels, and data patterns. Error bars show the range of \gls{ber} across rows.}
    \label{fig:berplot}
\end{figure}

\figref{fig:berplot} shows a box-and-whiskers plot \omcr{2}{depicting} the distribution of \gls{ber} (y-axis) across different DRAM rows for a given data pattern (x-axis) in a channel (color-coded).\footnote{\label{fn:boxplot}{\agycr{1}{The lower- and upper-bounds of the box are the first and the third quartiles, marking the medians of the first and second half  of the ordered set of data points, respectively.
Whiskers show the minimum and maximum values. The circle marker in each box shows the distribution's mean.}}} \agycr{2}{\agycr{3}{In addition to the four data patterns,} we choose \agycr{3}{and plot the \gls{ber} for} the \gls{wcdp} \agycr{3}{of} each row.}

We make {four} major observations from \figref{fig:berplot}.
First, {RH} bitflips \omcr{2}{occur in \emph{every} tested DRAM row across \emph{all}} HBM channels.
Second, \gls{ber} varies across channels. A subset of channels (channels {6} and {7}) exhibit significantly {higher \gls{ber}} than other channels. For example, channel 7 (with the highest \gls{ber}) has $\param{2.03}\times$ higher \gls{ber} than channel 0 (with the lowest \gls{ber}) for \gls{wcdp}. 
Third, channels can be classified into groups of two based on the number of bitflips they exhibit. We highlight {these groups using different shades of the same color in the figure}. We hypothesize that groups of channels are spread across \agycr{2}{\emph{different}} HBM2 DRAM dies. The \agycr{1}{variation} in \gls{ber} across the \agycr{1}{channel} groups could be due to \agycr{1}{manufacturing} process variation (similar to \agycr{1}{that in} DDR3/4 \agycr{1}{DRAM} chips~\cite{kim2014flipping,kim2020revisiting,orosa2021deeper, park2016statistical,park2016experiments}). {Fourth, \gls{ber} changes with data pattern, e.g., the maximum \gls{ber} in channel 7 is \param{3.13\%} and \param{2.04\%} for data patterns Rowstripe1 and Checkered0, respectively.}

\begin{figure}[ht]
    \centering
    \includegraphics[width=\linewidth]{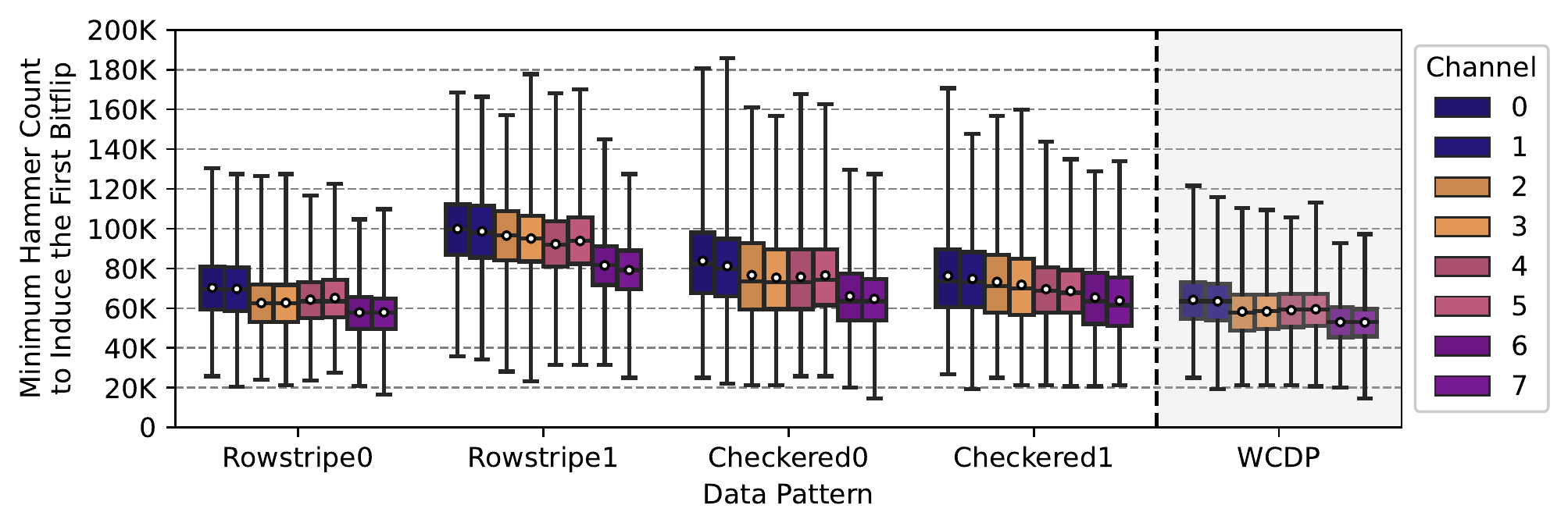}
    \vspace{-1em}
    \caption{\Glsfirst{hcfirst} across different DRAM rows, channels, and data patterns.}
    \label{fig:hcfplot}
\end{figure}

\figref{fig:hcfplot} shows the distribution of \gls{hcfirst} (y-axis) across different DRAM rows for a given data pattern (x-axis) in a channel (color-coded) \agycr{1}{using a box-and-whiskers plot}.{\fnref{fn:boxplot}} \agycr{2}{\agycr{3}{In addition to the four data patterns,} we choose \agycr{3}{and plot the \gls{hcfirst} for} the \gls{wcdp} \agycr{3}{of} each row.}

We make \param{three} major observations from \figref{fig:hcfplot}. First, \nrh{} is as low as \param{14531} across all tested channels and data patterns. Second, different channels exhibit different \nrh{} distributions. For example, \omcr{2}{channels} \param{7} and \param{6} contain more rows with smaller \nrh{} values than other channels. Because these channels also exhibit a higher number of RH bitflips than other channels, we hypothesize that these channels belong in the die with the worst RH vulnerability across all dies. Third, the \nrh{} distribution in a channel depends on the data pattern used. {For example, the mean \gls{hcfirst} for Rowstripe0 and Rowstripe1 in channel 0 are \param{57925} and \param{79179}, respectively.} {We conclude that testing with different data patterns is necessary to assess the RH vulnerability of HBM2 DRAM \agycr{2}{chips} as no data pattern achieves the smallest \gls{hcfirst} or \gls{ber} (\figref{fig:berplot}).}

\begin{figure}[ht]
    \centering
    \includegraphics[width=\linewidth]{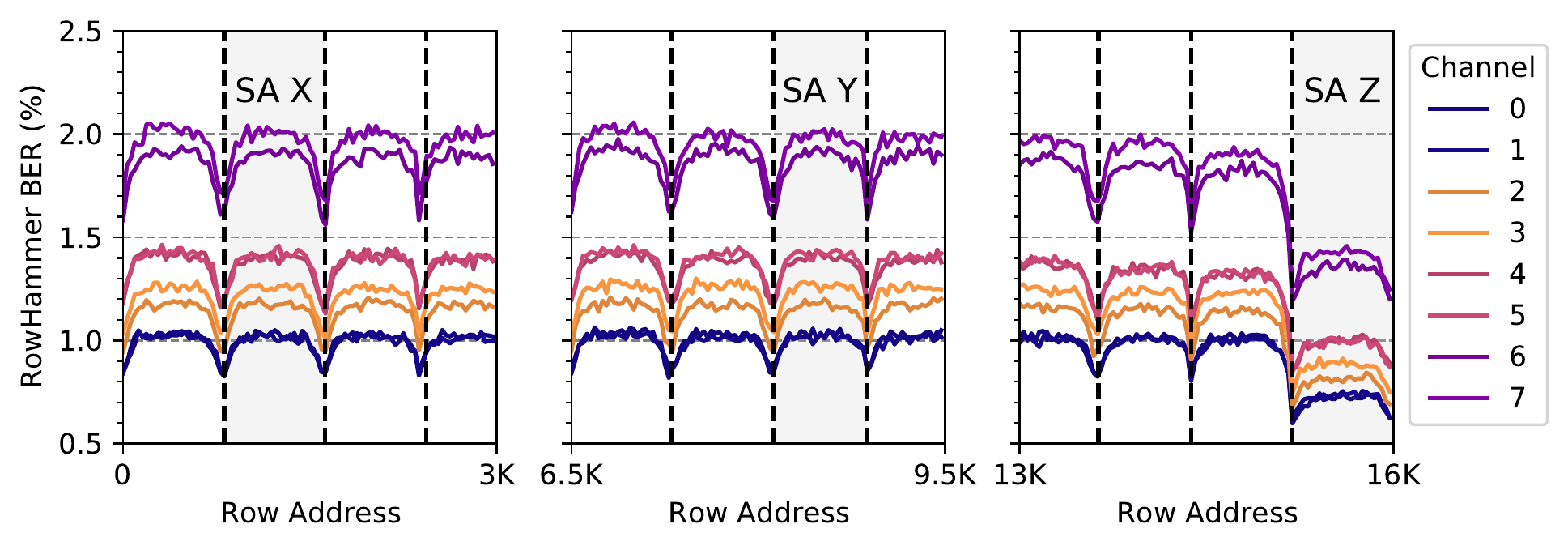}
    \vspace{-1em}
    \caption{\Glsfirst{ber} for different rows across a bank in different channels. Higlighted regions show individual DRAM {subarrays}.}
    \vspace{-3mm}
    \label{fig:beracrossrowsplot}
\end{figure}

\figref{fig:beracrossrowsplot} shows the BER (y-axis) for tested DRAM rows (x-axis) when we use \agycr{2}{the per-row \gls{wcdp}} to initialize the rows. Different channels are color-coded and different subplots show the three regions (first, middle, and last 3K rows) we test.

We make \param{two} major observations from \figref{fig:beracrossrowsplot}. \agycr{1}{First,} the \gls{ber} periodically increases and decreases across DRAM rows \omcr{2}{such that it} is higher in the middle of a \emph{subarray} and lower towards either end of the {subarray}.\footnote{{We reverse engineer subarray boundaries by performing single-sided RH~\cite{kim2014flipping,kim2020revisiting} that induces bitflips in \emph{only one} of the victim rows if the aggressor row is at the edge of a subarray. We find that a subarray contains either 832 (SA X in \figref{fig:beracrossrowsplot}) or \param{768} (SA Y in \figref{fig:beracrossrowsplot}) DRAM rows.}} {We hypothesize that this pattern results from the \agycr{2}{physical organization of the DRAM banks}. For example, the RH vulnerability of a row could increase with the row's distance from the row buffer. Second, \agycr{1}{significantly fewer bitflips occur in} the last subarray of the bank (\agycr{1}{e.g.,} the last 832 rows in SA Z) \agycr{1}{compared to the rest of the bank.}
We hypothesize that 
\agycr{1}{\omcr{2}{this} is} due to the \agycr{1}{DRAM bank's physical placement.}
For example, assuming that proximity to the shared I/O circuitry on the DRAM die affects the RH vulnerability of a subarray, the last subarray might be placed near this shared I/O circuitry~\cite{jun2017hbm}.
\agycr{1}{We leave further analysis on subarrays \omcr{2}{to} future work.}
}

\begin{figure}[!h]
    \centering
    \includegraphics[width=\linewidth]{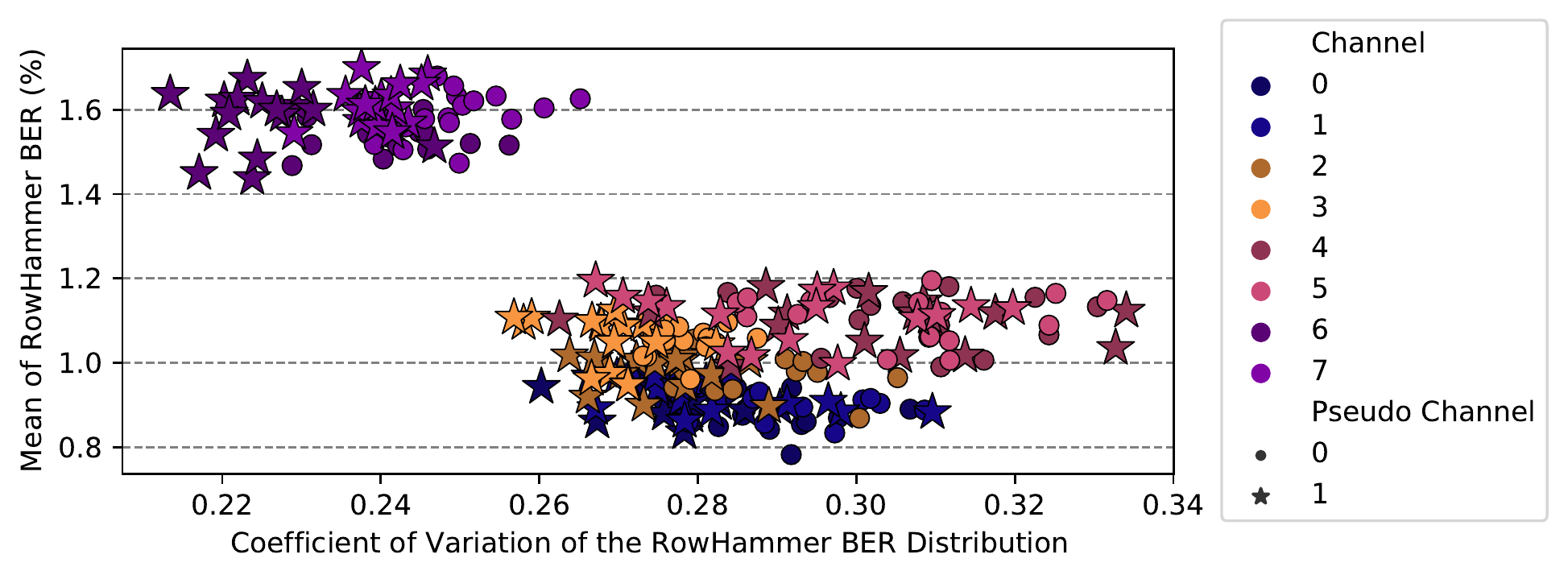}
    \vspace{-1em}
    \caption{\gls{ber} variation across banks. Each bank is represented by the average BER (y-axis) and the coefficient of variation in BER (x-axis) across the rows within the bank.}
    \vspace{-2mm}
    \label{fig:bankvar}
\end{figure}

\agycr{2}{We examine RH vulnerability differences across banks by comparing the \stecr{2}{RH} \gls{ber} distribution over 300 rows}
\agycr{2}{(the first, the middle, and the last 100 rows) \omcr{3}{in each of the} 256 banks (\omcr{3}{across} 8 channels and two pseudo channels per channel). \figref{fig:bankvar} shows each bank in a scatter plot based on its \gls{ber} distribution: 1)~the \gls{cov}\footnote{\Gls{cov} is a distribution's standard deviation, normalized to its mean~\agycr{2}{\cite{brian1998dictionary}}.} on the x-axis and 2)~the mean on the y-axis. A marker's color and shape indicate the bank's channel and pseudo channel, respectively.}

We make \param{two} major observations. First, there is variation in \gls{ber} across banks and pseudo channels. For example, there is up to \param{0.23}\% difference in mean \gls{ber} across banks in channel \param{7}. Second, \gls{ber} variation across banks is dominated by variation across channels. Banks in different channels tend to have a larger \gls{ber} difference than banks in the same channel (\figref{fig:berplot}). We conclude that testing different channels is more important than testing different banks or pseudo channels \agycr{1}{to assess} the RH vulnerability of a given HBM2 DRAM chip.

\noindent
\textbf{\agycr{1}{Summary}.} We summarize the two key takeaways of our spatial variation study. First, different 3D-stacked channels of the HBM2 DRAM chip exhibit significantly different levels of {RH} vulnerability in terms of \gls{ber} (up to \SI{79}{\percent}) and \gls{hcfirst} (up to \SI{20}{\percent}). Second, DRAM rows
near the end of a DRAM bank exhibit substantially smaller \gls{ber} than other DRAM rows on average across all HBM2 DRAM channels. {Our key takeaways have two implications for RH attacks and defenses. First, an RH attack can use the most-RH-vulnerable HBM2 channel to reduce the time it spends on i) \emph{preparing} for an attack, by finding exploitable RH bitflips faster (i.e., by accelerating memory templating), and ii) \emph{performing} the attack, by benefiting from a small \gls{hcfirst} value. Second, an RH defense mechanism can adapt itself to the heterogeneous distribution of the RH vulnerability across channels and subarrays, which may allow the defense mechanism to more efficiently prevent RH bitflips.}

\section{Uncovering in-DRAM \agycr{1}{RH} Mitigations}
\label{sec:uncovering}

To prevent \agycr{1}{RH} bitflips, DRAM manufacturers equip their chips with a mitigation mechanism broadly referred to as Target Row Refresh (TRR)~\cite{hassan2021utrr,frigo2020trrespass,micron2016trr}. Proprietary versions of TRR (e.g., in DDR4) operate transparently from the perspective of the memory controller. At a high level, TRR identifies potential aggressor rows and preventively refreshes their victim rows~{upon receiving a $REF$ command}. 
We demonstrate that the tested chip implements a proprietary TRR (similar to the ones used in DDR4~\cite{hassan2021utrr, hassan2021utrrgithub}) \agycr{1}{in addition to} the TRR mode \omcr{2}{documented in the HBM2 standard}~\cite{jedec2015hbm}. 

\noindent
\textbf{Methodology.}
{
We use \omcr{2}{the U-TRR} methodology~\cite{hassan2021utrr, hassan2021utrrgithub} to uncover the proprietary TRR mechanism. {The key idea of this methodology is to use retention failures as a side channel to infer whether or not TRR refreshes a DRAM row. 

One iteration of our experiment consists of six steps. First, we profile a row (R) to find its retention time (T), after which retention errors accumulate in row R unless row R is refreshed. Second, we activate and precharge row R once (i.e., refresh \omcr{2}{it}) and wait for T/2. Third, we activate and precharge row R+1. We hypothesize that if the HBM2 chip implements an undisclosed TRR mechanism, it will sample the activation of R+1 and refresh the neighboring rows (i.e., row R) once a periodic $REF$ command triggers \omcr{2}{TRR}. Fourth, to trigger the TRR mechanism, we issue a periodic $REF$ command. Fifth, we wait for another T/2 such that if TRR does \emph{not} refresh row R, it will exhibit retention errors. Sixth, we check row R for bitflips to see if \emph{any} TRR refresh \omcr{2}{occurred (i.e., if there are \emph{no} bitflips)}. We \agycr{2}{perform} 100 \agycr{2}{iterations of this experiment to check} if TRR refreshes the victim row R.}}

\noindent
\textbf{Results.}
{We observe that the profiled row (R) is refreshed \emph{once} every \emph{17} iterations. Therefore, we conclude that i) the tested HBM2 DRAM chip implements a proprietary, undisclosed TRR mechanism, and ii) this TRR mechanism performs a victim row refresh once every 17 periodic $REF$ commands, {resembling} \agycr{2}{a} TRR mechanism that U-TRR~\cite{hassan2021utrr, hassan2021utrrgithub} uncovers in DRAM chips from \emph{Vendor C}. We \omcr{2}{intend to} uncover more details of \agycr{2}{the} proprietary TRR mechanism as part of future work.}

%% file: sections/06_future_work.tex
\section{Future Work}
{This work presents the results of our \stecr{2}{RH} characterization study on a real HBM2 DRAM chip.}
We plan to {present more insights into how \stecr{2}{RH} behaves in real HBM2 DRAM chips by strengthening our characterization study at least in the following \omcr{3}{three} directions}.

\noindent
\textbf{1) Testing more HBM chips}. We \omcr{2}{intend to} repeat our experiments on a large{r} number of {HBM2} {chips} {to} {improv{e} the statistical significance of our observations}. 

\noindent
\textbf{{2) More characterization} for each HBM chip}. {W}e plan to understand \stecr{2}{RH}'s sensitivities to multiple other factors. We will investigate how {RH varies: 1) across different stacks in HBM2 chips, 2) based on the time an aggressor row remains active~\cite{orosa2021deeper}, 3) based on a richer set of data patterns used in initializing victim and aggressor rows, and 4) across different HBM2 voltage and temperature levels.} \omcr{2}{We also intend to study the effect of RowPress~\cite{luo2023rowpress}.}

\noindent
\textbf{3) Investigating cross-channel interference.} HBM2 chips stack DRAM dies such that certain HBM channels are placed on top of each other. We \omcr{2}{aim to} investigate if frequently accessing one or more \emph{aggressor channels} can induce bitflips or worsen the reliability characteristics of other \emph{victim channels}.\omcrcomment{2}{Do we want to give out this particular direction? Giray and Ataberk: We are fine with that.}

%% file: sections/07_related_work.tex
\section{Related Work}

{We present the first experimental characterization of the \omcr{2}{RowHammer} \stecr{2}{(RH)} vulnerability in a modern HBM2 DRAM chip. Prior works~\understandingRowHammerAllCitations{} analyze new aspects of the RH vulnerability by testing real DDR3/4 DRAM chips. Other prior works~\cite{larimi2021understanding, kwon2023temperature,sullivan2021characterizing} characterize real HBM chips to understand their i)~soft error resiliency~\cite{sullivan2021characterizing}, ii)~performance and reliability characteristics under reduced voltage~\cite{larimi2021understanding}, and iii)~susceptibility to retention failures at different temperatures~\cite{kwon2023temperature}. \omcr{2}{No prior work investigates} the RowHammer vulnerability in a real HBM chip.}

%% file: sections/08_conclusion.tex
\section{Conclusion}

\omcr{2}{We} present the results of our detailed characterization study \omcr{2}{of} the RowHammer \stecr{2}{(RH)} vulnerability in a modern HBM2 chip. We show that the RH vulnerability is heterogeneously distributed across various components in the HBM2 chip, which \omcr{2}{we believe} has important implications for future RH attacks and defenses. We discover that the HBM2 chip implements a proprietary RH mitigation mechanism and explain how the mitigation mechanism works. \agycr{2}{We hope and expect that our findings will lead to a deeper understanding of and new solutions to the RH vulnerability in HBM-based systems.}